\documentclass[preprint,showpacs,preprintnumbers,amsmath,amssymb]{revtex4}
\usepackage{graphicx}
\usepackage{dcolumn}
\usepackage{bm}

\begin{document}

\title{Pairing and the phase diagram of the normal coherence length $\xi_N(T,x)$ above $T_c$
of $La_{2-x}Sr_xCuO_4$ thin films\\ probed by the Josephson effect}

\author{Tal Kirzhner}

\affiliation{Physics Department, Technion-Israel Institute of Technology,
Haifa 32000, Israel}

\author{Gad Koren}

\email{gkoren@physics.technion.ac.il} \affiliation{Physics Department, Technion-Israel Institute of Technology, Haifa 32000, Israel} \homepage{http://physics.technion.ac.il/~gkoren}

\date{\today}

\begin{abstract}

The long range proximity effect in high-$T_c$ c-axis Josephson junctions with a high-$T_c$ barrier of lower $T_c$ is still a puzzling phenomenon. It leads to supercurrents in junctions with much thicker barriers than would be allowed by the conventional proximity effect. Here we measured the $T-x$ (Temperature-doping level) phase diagram of the barrier coherence length $\xi_N(T,x)$, and found an enhancement of $\xi_N$ at moderate under-doping and high temperatures. This indicates that a possible origin of the long range proximity effect in the cuprate barrier is the conjectured pre-formed pairs in the pseudogap regime, which increase the length scale over which superconducting correlations survive in the seemingly normal barrier. In more details, we measured the supercurrents $I_c$ of Superconducting - Normal - Superconducting SNS c-axis junctions, where S was optimally doped $YBa_2Cu_3O_{7-\delta}$ below $T_c$ (90 K) and N was $La_{2-x}Sr_xCuO_4$ above its $T_c$ ($<$25 K) but in the pseudogap regime. From the exponential decay of $I_c(T)\propto exp[-d/\xi_N(T)]$, where $d$ is the barrier thickness, the $\xi_N(T)$ values were extracted. By repeating these measurements for different barrier doping levels $x$, the whole phase diagram of $\xi_N(T,x)$ was obtained.

\end{abstract}

\maketitle

A controversy still exist concerning the nature of the pseudogap regime in the cuprate superconductors \cite{Timusk,Fischer}. Some researchers visualize the pseudogap regime as a precursor to superconductivity, where uncorrelated  pairs which form below the pseudogap cross-over temperature $T^*$, acquire global phase coherence at $T_c$ \cite{Emery}. Others view the pseudogap regime as a phase or phases which are competing with superconductivity such as in spin and charge density waves and when charge, magnetic and gyrotropic orders occur \cite{Guy,TanakaK,Blanco,LeBoeuf,Fauque,Xia,Hosur,YuvalL,Varma}. The former group bases its case mostly on tunneling and ARPES measurements  \cite{Fischer,Kanigel,MingShi} of a single energy gap which evolves smoothly while crossing from the superconducting phase into the pseudogap regime. The latter group uses different observations of two distinct energy gaps, obtained by the same techniques, to rest their case \cite{Shen,Boyer}. The whole picture of the pseudogap regime however, seems to be much more complex as various experiments show precursor superconductivity coexisting with competing orders in the same samples \cite{Blanco,Dubroka,Wu}. A possible origin for the competing and coexisting orders is the inherent inhomogeneity of the surface of the cuprates, but global measurements which average over these inhomogeneities, still bring up new results which lend support to one or more of the above mentioned scenarios \cite{Dong}. Polarized elastic neutron scattering and  ultrasound measurements in $YBa_2Cu_3O_{6+x}$ have shown that the pseudogap is bound by a line of a real thermodynamic phase transition rather than by a cross over regime only \cite{Fauque,Shekhter}. So the controversy on the origin of the pseudogap regime is still ongoing \cite{PT2013}.\\

Here we focus on properties of the pseudogap as revealed by supercurrent measurements in superconducting - normal - superconducting SNS Josephson junctions, where N is in the pseudogap regime of a cuprate barrier with a $T_c$ lower than that of S. The observed results are closely related to the long-range (or "giant") proximity effect in trilayer c-axis junctions which was investigated previously both experimentally and theoretically  \cite{Bozovic,Bergeal,Covaci,Marchand}. In one of these studies, supercurrents were observed also at temperatures significantly above $T_c$ of the N-barrier, even when its thickness was two orders of magnitude larger than the expected "normal" coherence length $\xi_N\approx 0.2$ nm for transport in the c-axis direction \cite{Bozovic}. The actual $\xi_N$ is therefore long ranged compared to that of the conventional proximity effect, and seems to reflect the specific nature of the pseudogap regime with its conjectured preformed pairs. To further substantiate this hypothesis, a systematic study of the supercurrent $I_c$ dependence on temperature $T$ and barrier doping level $x$ is needed. This was done in the present study using $YBa_2Cu_3O_{7-\delta}- La_{2-x}Sr_xCuO_4-YBa_2Cu_3O_{7-\delta}$ junctions (for $x$=0.07, 0.1, 0.18 and 0.24), with the intention of obtaining a phase diagram of $\xi_N(T,x)$ from the measured $I_c(T,x)$ data. Since the proportionality constant of the proximity relation $I_c \propto exp[-d/\xi_N]$ is unknown, we had to have $I_c(T,x)$ data for at least two different $d$ values for each doping level $x$ in order to extract $\xi_N(T,x)$. Once done, we present a novel phase diagram of $\xi_N(T,x)$ of $La_{2-x}Sr_xCuO_4$ above its $T_c$ and in its pseudogap regime, where on the $T$ versus $x$ diagram, the contours of constant $\xi_N(T,x)$ for $T>55$ K have a maximum in the underdoped regime. This provides further supporting evidence for the precursor superconductivity scenario in the cuprates.\\

\begin{figure}
\includegraphics[height=6.5cm]{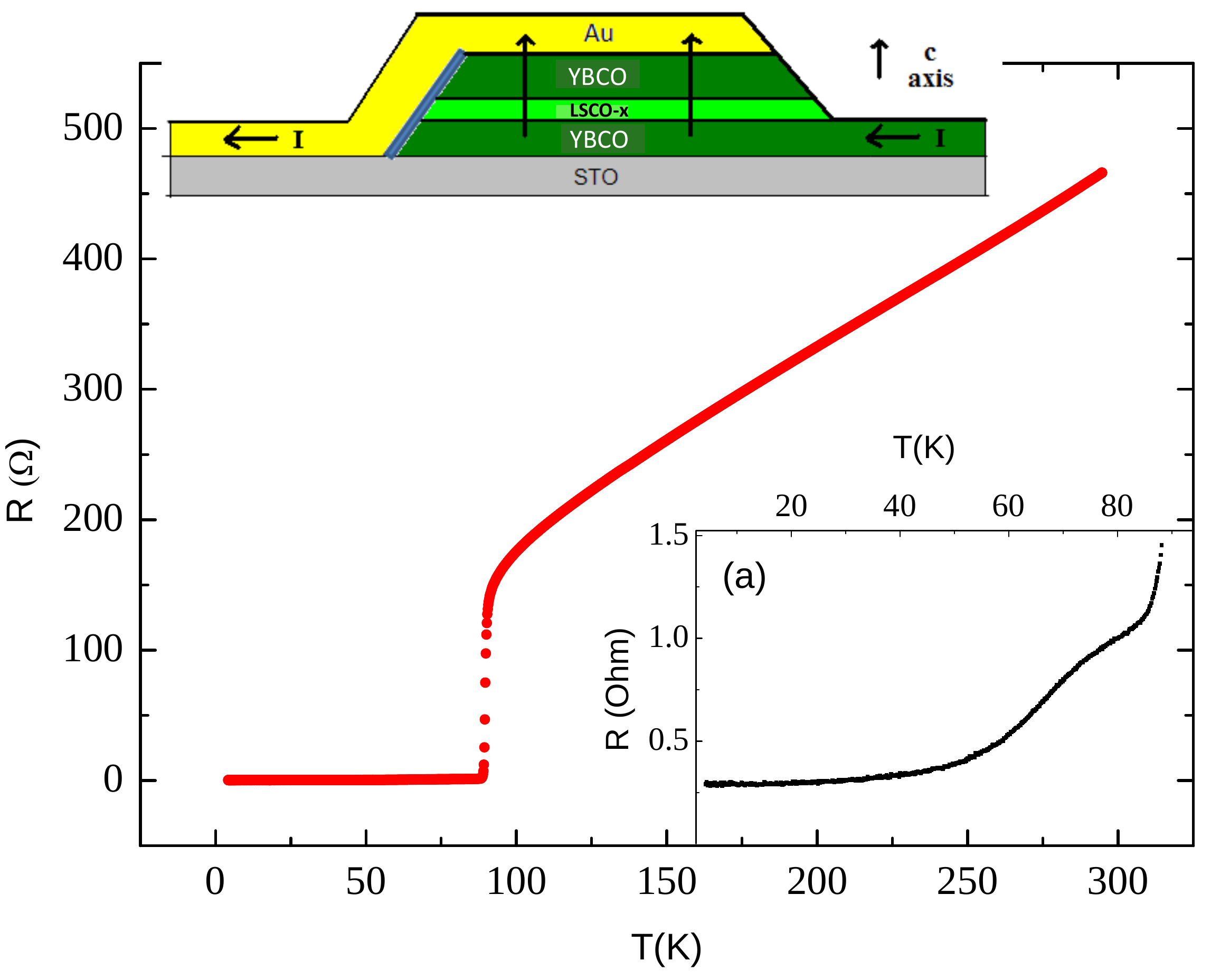}
\caption{\label{fig:RT} Resistance versus temperature of a Josephson junction with a 20 nm thick LSCO-0.07 barrier obtained using a 100 $\mu$A current bias. The top inset shows a schematic cross-section of the junction, and the bottom one is a zoom-in on the resistance below the transition of the YBCO electrodes. }
\end{figure}

\section{Preparation of the junctions}
We chose to work with fully epitaxial SS'S thin film junctions of the cuprates that have a conveniently wide temperature range where S' is in the pseudogap regime between the $T_c$ values of S' and S. In this regime, we shall refer to the junctions as SNS junctions, which is the more commonly used term in such a situation. Optimally doped $YBa_2Cu_3O_{7-\delta}$ (YBCO) with $T_c\approx 90$ K was chosen as the S electrodes, while the S' barrier was chosen to be $La_{2-x}Sr_xCuO_4$ (LSCO-x) with $T_c$ values of up to about 25 K. A schematic cross-section  of a junction is shown in the top inset of Fig. 1. The trilayer film of YBCO/LSCO-x/YBCO was grown epitaxially in-situ by laser ablation deposition on $10\times 10$ mm$^2$ wafers of (100) $SrTiO_3$. The trilayer was then patterned by photolithography and Ar ion milling to produce ten base electrodes with their corresponding ramps on the wafer. This was followed by a room temperature deposition of the gold cover electrode, which unlike in our previous ramp junctions \cite{TalCARE}, left the ramp of the base electrode in a highly resistive state, with only a negligible current flow in the a-b plane direction through it for the lack of the high temperature annealing step. This yielded a cross-over junction where the current flows mostly in the c-axis direction via a $5\times 5\,\mu m^2$ area (defined by a second patterning process) into the gold cover electrode.\\

\begin{figure}
\includegraphics[height=6.5cm]{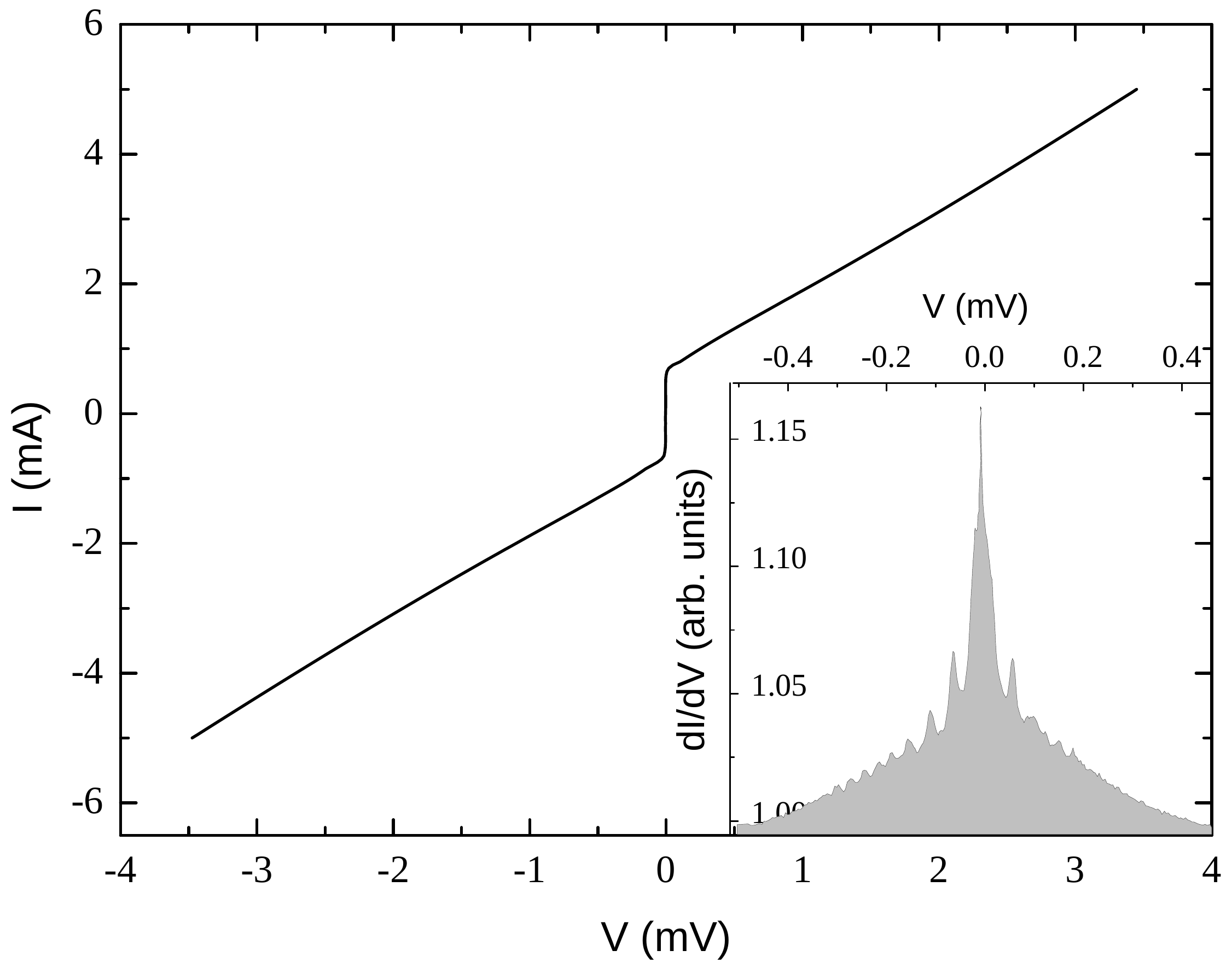}
\caption{\label{fig:IV} Current versus voltage at 10 K of a Josephson junction with a 20 nm thick LSCO-0.07 barrier. A gold series resistance of 0.3 $\Omega$ was subtracted from the data. The inset shows the conductance of this junction at 40 K under 10.7 GHz microwave irradiation, where the Shapiro steps in the corresponding I-V curve are seen as evenly spaced peaks. }
\end{figure}

\section{Transport results}
Fig. \ref{fig:RT} shows a typical resistance versus temperature curve of a Josephson junction with a 20 nm thick $La_{1.93}Sr_{0.07}CuO_4$ barrier with 7\% Sr doping (LSCO-0.07). The YBCO electrodes become superconducting at $T_c\approx$90 K where the junction resistance drops sharply. It doesn't however drop to zero, as can be seen in the bottom inset of Fig. \ref{fig:RT} which shows a knee-like structure down to about 50 K on top of an almost constant residual resistance of $\sim 0.3\,\Omega$ below it. These two resistance components originate in the barrier material in the junction (the LSCO-0.07 layer) and the gold cover electrode. Once the LSCO-0.07 layer becomes superconducting at about 50 K by the proximity effect, the constant residual resistance below it is due to the gold cover electrode only. Thus, as we lower the temperature further, the Josephson current increases but the series resistance of the Au cover electrode remains.\\

A typical I-V curve at 10 K of this type of Josephson junction is shown in Fig. \ref{fig:IV}. This curve shows that the junction has a critical current of 0.55 mA as measured by a 5 $\mu V$ criterion. It also exhibits a resistively shunted junction (RSJ) behavior at higher bias with a normal resistance of $0.8\,\Omega$. The $I_cR_N$ product of the junction is therefore equal to 0.44 meV  which is typical of Josephson junctions in the cuprates ~\cite{SNS}. The inset of Fig. \ref{fig:IV} depicts a conductance spectrum of this Josephson junction at 40 K under 10.7 GHz microwave irradiation, showing the AC Josephson effect. The evenly spaced peaks in the curve are due to Shapiro steps in the I-V curve at a somewhat larger than the expected spacing of $\Delta V=h\nu/2e$ due to the series resistance of about 0.3 $\Omega$ of the gold cover electrode. On a wafer with 10 junctions, the spread of the measured critical current values was about $\pm$30\%. In the following measurements of $I_c$ versus T on each wafer with a given barrier thickness and doping level, we had chosen to work on the junction whose critical current value is closest to the average value obtained on that wafer.\\

\begin{figure}
\includegraphics[height=6.5cm]{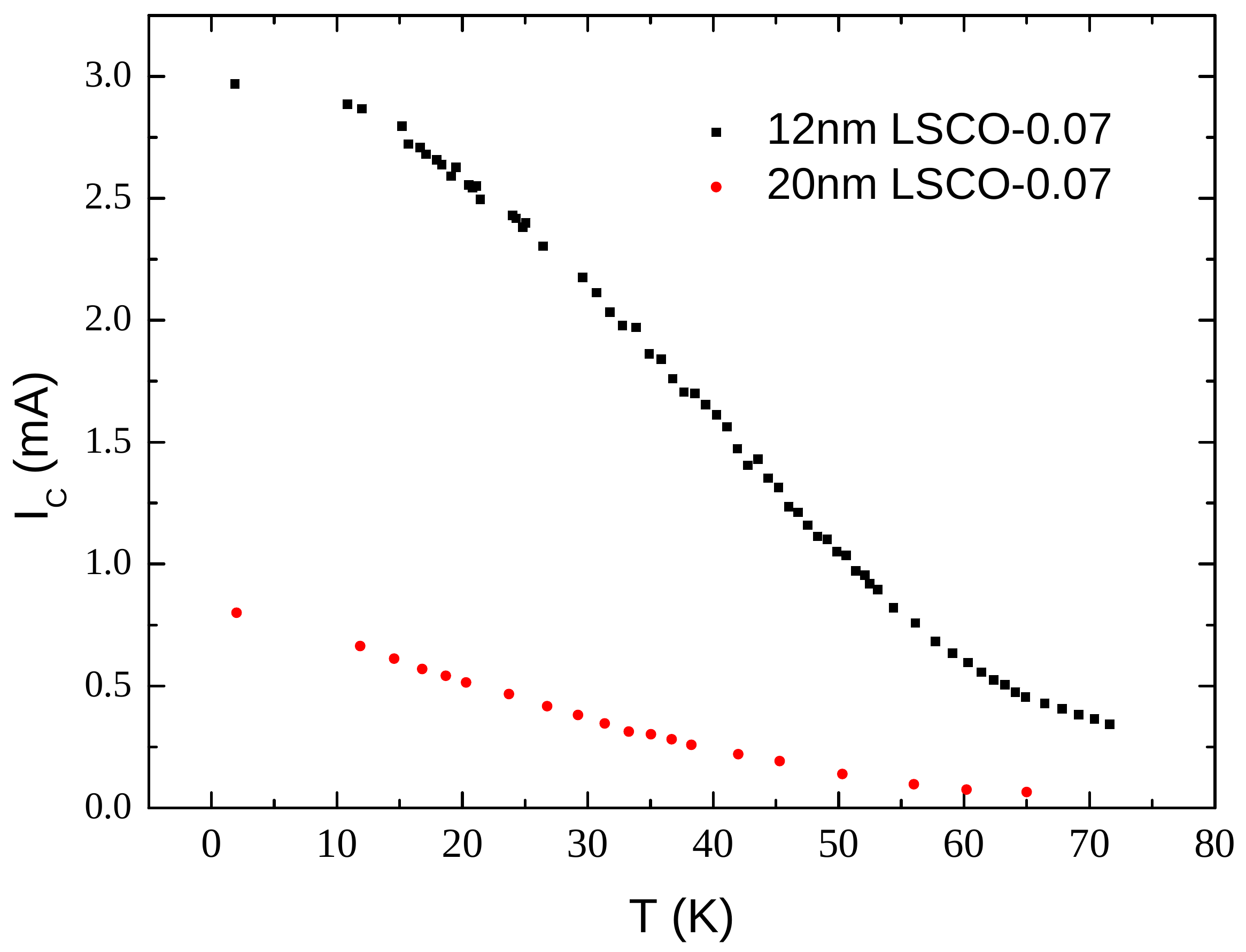}
\caption{\label{fig:Temp} Critical current as a function of temperature of two Josephson junctions on two different wafers with LSCO-0.07 barrier thicknesses of 20 nm and 12 nm. }
\end{figure}

Next we focus on the temperature dependence of the critical currents which were extracted from the I-V curves using a 5 $\mu V$ criterion. Fig. \ref{fig:Temp} shows the temperature dependence of the critical currents in two representative Josephson junctions with LSCO-0.07 barriers on two different wafers. One with a 20 nm thick barrier and the other with a thinner 12 nm thick barrier. At temperatures above 40-50 K, when the critical currents are small, they decay versus temperature as a $(T_c-T)^2$ power law as predicted by the De-Gennes dirty limit proximity effect formula \cite{DG}. At lower temperatures, as the supercurrents increase, the I-V curves deviate from the weak-link RSJ model and start to show a strong-link flux flow behavior which changes the power law temperature dependence. This may be due to the self field effect when the width of the junction $w$ becomes larger than the Josephson penetration depth $\lambda_J$. A critical current of about 1.5 mA at 40 K with the 12 nm thick barrier, corresponds to a Josephson penetration depth $\lambda _J$ of $ \simeq 4 \,\mu m$ which is of the order of the width of our junctions ($w=5\,\mu$m).\\

\section{Extraction of $\xi_N(T,x)$ from the data}
\begin{figure}
\includegraphics[height=6.5cm]{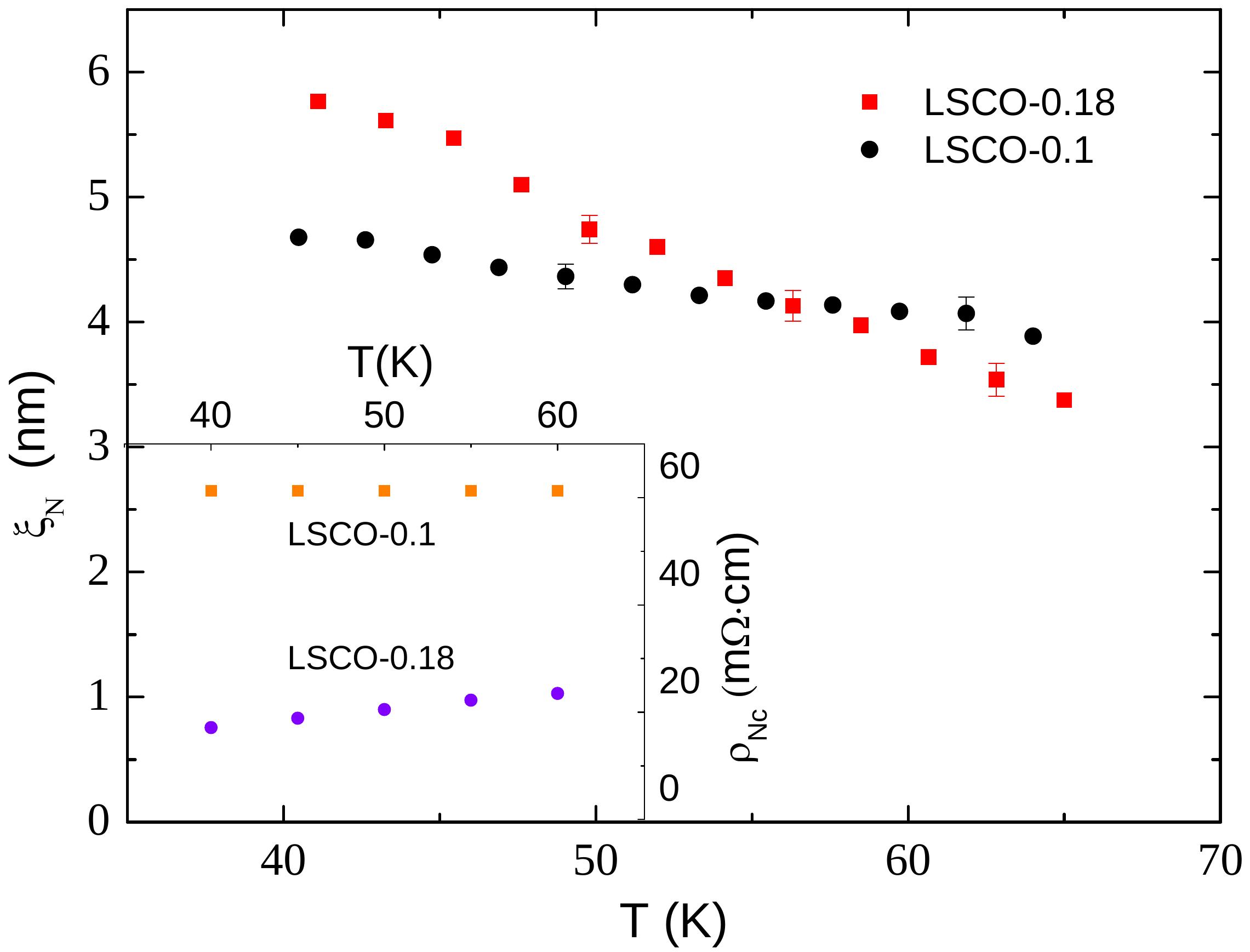}
\caption{\label{fig:coherence} Normal coherence lengths $\xi_N$ of LSCO-0.1 and LSCO-0.18 as a function of temperature. The inset shows the normal state resistivity of the corresponding LSCO-x barriers as a function of temperature. }
\end{figure}

We now turn to the main result of this study which shows the normal coherence lengths of the LSCO-$x$ barriers at different doping levels. For any given temperature $T$ and doping level $x$, the normal coherence length of the barrier can be extracted from the ratio of the critical currents in junctions with two different barrier thicknesses $d_i$ using the exponential part of the De-Gennes formula ($I_{ci}\propto exp[-d_i/\xi_N]$ \cite{DG}). To further clarify the procedure of extracting $\xi_N(T)$ from the data, a detailed description is given in the supplementary material for the case of LSCO-0.24 film and junctions \cite{supplementary}. Fig. \ref{fig:coherence} shows the normal coherence lengths $\xi_N(T,x)$ for $x$=0.1 and $x$=0.18 LSCO-x barriers as a function of temperature. The temperature range of the coherence plots is limited here to 40-60 K. The lower bound of the temperature range is set by the flux flow phenomenon in the junctions with the thinner barrier due to the high $I_c$ values and rounding of the I-V curves which make the determination of $I_c$ difficult. The upper bound is set by the low critical currents in the junctions with the thicker barrier which are noisy and therefore hard to measure. Fig. \ref{fig:coherence} shows that the measured normal coherence length values range between 4-6 nm. These values are much higher than expected from the conventional proximity effect theory \cite{DG}, where the coherence length should be limited by the short c-axis superconductor coherence length $\xi_S$ and the corresponding mean free path $l_N$, both of which are shorter than 1 nm. Previous experiments on SNS cuprate junctions of the type LSCO-LCO-LSCO had also shown very long coherence lengths \cite{Bozovic}. This "giant proximity effect" was explained by a number of theories which took into account superconducting phase fluctuations above $T_c$ in the barrier \cite{Covaci,Marchand}.\\

\section{Discussion}
Another interesting feature in Fig. \ref{fig:coherence} is the unexpected crossing of the $\xi_N$ curves at about 55 K for the two doping levels. At low temperatures, the overdoped LSCO-0.18 barrier has a higher normal coherence length than the underdoped LSCO-0.1. This behavior however is reversed above 55 K, where the coherence length of the LSCO-0.1 barrier becomes higher. In the following we shall try to understand this peculiar dependence of $\xi_N$ of LSCO-x which is a dirty limit material for transport in the c-axis direction. As we shall calculate only the ratio of coherence lengths $\xi_{N}(0.18)/\xi_{N}(0.1)$, any effect of the long range proximity effect should cancel out to a first approximation. Moreover, long range proximity effect was obtained using the standard proximity effect while invoking strong superconducting "pockets" in the barrier \cite{Covaci}, which is similar to the pre-formed pairs scenario \cite{Emery}. We shall thus use here the conventional proximity effect theory. In the dirty limit this yields the normal coherence length

\begin{equation}\xi_{Nd}=\sqrt{\frac{\hbar D_N}{2\pi k_BT}}
\end{equation}
where $D_N$ is the diffusion constant. In order to estimate the value of $D_N$, we used the normal resistivity values obtained from the I-V curves of our junctions at high bias. The resulting resistivity values $\rho_{Nc}(T,x)$ are plotted in the inset of Fig. \ref{fig:coherence}. In our junctions the current flows in the c-axis direction and therefore these $\rho_{Nc}(T,x)$ results represent inter-layer transport. In the anisotropic cuprates, a prevalent model for the transport mechanisms assumes a strong in-plane coupling where superconductivity occurs, and a weak-link, Josephson coupling between the planes. One such model was discussed by Graf, Rainer and Sauls \cite{Graf}, where the normal c-axis, inter-layer conductivity $\sigma_{Nc}=1/\rho_{Nc}$ is given by:
\begin{equation}\sigma _{Nc}=2N_fe^2D
\end{equation} where $N_f$ is the density of states at the chemical potential.
This allowed us to calculate the inter-layer diffusion constant $D$ and the corresponding $\xi_{Nd}$ of Eq. (1), using the measured resistivity values while the density of states values were taken from Ino et al \cite{DOS}. Using this procedure, we calculated the ratio between the normal coherence lengths of the $x$=0.18 and $x$=0.1 Sr doped barriers.  This yields $\frac{\xi_{Nd}(0.18)}{\xi_{Nd}(0.1)}\simeq 1.2\pm 0.1$ at 60 K, which disagrees with the measured coherence lengths of Fig. \ref{fig:coherence} at this temperature. Generally, the diffusion constant $D$ and therefore also $\xi_N$, should be larger in the less resistive materials, those with the higher doping level, as is actually seen in the low temperature regime of Fig. \ref{fig:coherence}. The fact that this behavior is reversed at the high temperature regime, must be due to an unconventional proximity effect where the LSCO-0.1 barrier does not behave as a normal metal. Some feature of this barrier should facilitate the long range proximity effect and the preference for higher $\xi_N$ values in the underdoped regime at higher temperatures. We attribute this behavior to the  precursor superconductivity scenario, in which the conjectured uncorrelated pairs (preformed pairs) allow for these phenomena to occur.\\

\begin{figure}
\includegraphics[height=6.5cm]{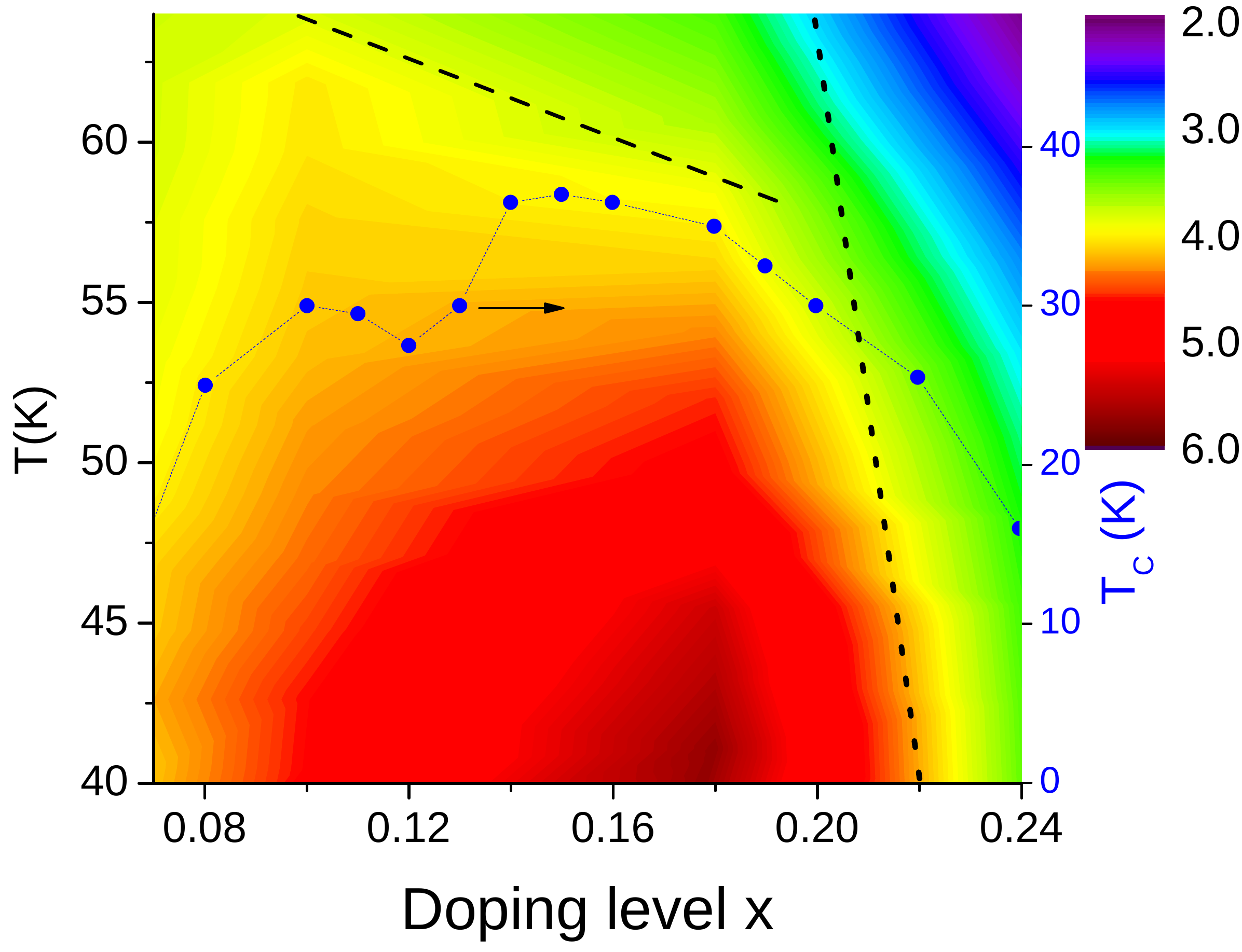}
\caption{\label{fig:colormap} A color-map of the phase diagram of $\xi_N(T,x)$ representing the normal coherence length of LSCO-$x$ in nm as function of  temperature $T$ and doping $x$. The dotted line represents the pseudogap T$^*$ temperature of Ref. \cite{Yoshida}, while the dashed line describes the trend of the present data of $\xi_N$ for $0.18\geq x \geq 0.1$ and $T>55$ K. For comparison we plot also $T_c(x)$ measured on LSCO single crystals by T. Matsuzaki, Phys. Chem. Sol. \textbf{62}, 29 (2001).  }
\end{figure}

\section{The phase diagram of $\xi_N(T,x)$ }
To further elucidate and explain this interpretation of our results, we plot in Fig. \ref{fig:colormap} a color-map of the full phase diagram of $\xi_N(T,x)$. All the measured $\xi_N(T)$ values of the $x$=0.07, 0.1, 0.18 and 0.24 doping levels were used (12$\times$4 measured values at 12 temperatures per each doping level), and the color-map extrapolates and draws the contours in between these doping levels. The contours in between the measured data points should thus be considered only as guides to the eye. We have data also below 40 K, but this is less reliable due to flux flow effects and we have chosen not to show it here. A clear feature in Fig. \ref{fig:colormap} is that the contours of constant $\xi_N$ follow roughly the superconducting dome, but this occurs much above the $T_c$ values of the LSCO-x barrier. Moreover, above 55 K, the maximum $\xi_N$ values for each contour occurs at moderate under-doping ($x$=0.1). One can see this behavior also by looking at the dashed line which shows the general trend of the contours in the $0.1<x<0.24$ doping range at high temperatures. Although reminiscent of the pseudogap $T^*$ behavior as depicted from ARPES measurements by the dotted line \cite{Yoshida}, the slopes of the two lines are very different, possibly indicating the presence of additional effects such as phase fluctuations or that the two phenomena are unrelated \cite{Shekhter,PT2013}. Similar phase diagram trends were observed before in the cuprates in Nernst effect measurements \cite{Ong}, in high magnetic field results \cite{Rullier}, in infrared and terahertz spectroscopy \cite{Dubroka,Bilbro}, and in higher energy gap results obtained in Andreev conductance spectroscopy measurements \cite{Koren}. These previous results, as well as the new one presented here, provide additional support for strong superconducting fluctuation effects and the preformed pairs scenario in the underdoped regime of the cuprates above $T_c$, but not necessarily up to the $T^*$ transition-line of the pseudogap.\\

In conclusion, comparative supercurrent measurements in SNS  YBCO - LSCO-x - YBCO c-axis junctions at various temperatures and doping levels $x$, yielded a novel phase diagram of $\xi_N(T,x)$, which besides the observation of a long range proximity effect, also support the precursor superconductivity scenario in the underdoped regime of the cuprates above $T_c$.\\


\section{Acknowledgments}
 This research was supported in part by the Israel Science Foundation, the joint German-Israeli DIP project and the Karl Stoll Chair in advanced materials at the Technion.\\

\section{Authors contributions}
T. Kirzhner had initiated this project and made the major part of the experiments. G. Koren made some of the experiments and both have written this manuscript.\\

\newpage

\begin{center} SUPPLEMENTARY MATERIAL\\
\vspace{1cm} for\\
\vspace{1cm}
 Pairing and the phase diagram of the normal coherence length $\xi_N(T,x)$ above $T_c$\\
of $La_{2-x}Sr_xCuO_4$ thin films probed by the Josephson effect\\

\vspace{1cm}
Tal Kirzhner and Gad Koren\\

\end{center}

In this supplementary part we describe in details three subjects. The first is transport measurements involving overdoped LSCO-0.24 film and junctions, including the critical current measurements versus temperature ($I_c(T)$). The second is how the normal coherence length $\xi_N(T)$ is extracted from the $I_c(T)$ values of two different junctions on two different wafers, both with the same LSCO-0.24 as the barrier. The third is a demonstration of the standard exponential decay of $\xi_N$ with the barrier thickness as found for the conventional proximity effect. This is done for similar junctions, but with a different cuprate barrier that also has a lower $T_c$ than that of the YBCO electrodes. This result is presented here in order to refute a conjectured logarithmic decay as proposed in the literature to explain the long range proximity effect \cite{Marchand}.\\

\section{Transport measurements of the LSCO-0.24 film and junctions}

\begin{figure}
\includegraphics[height=9cm,width=13cm]{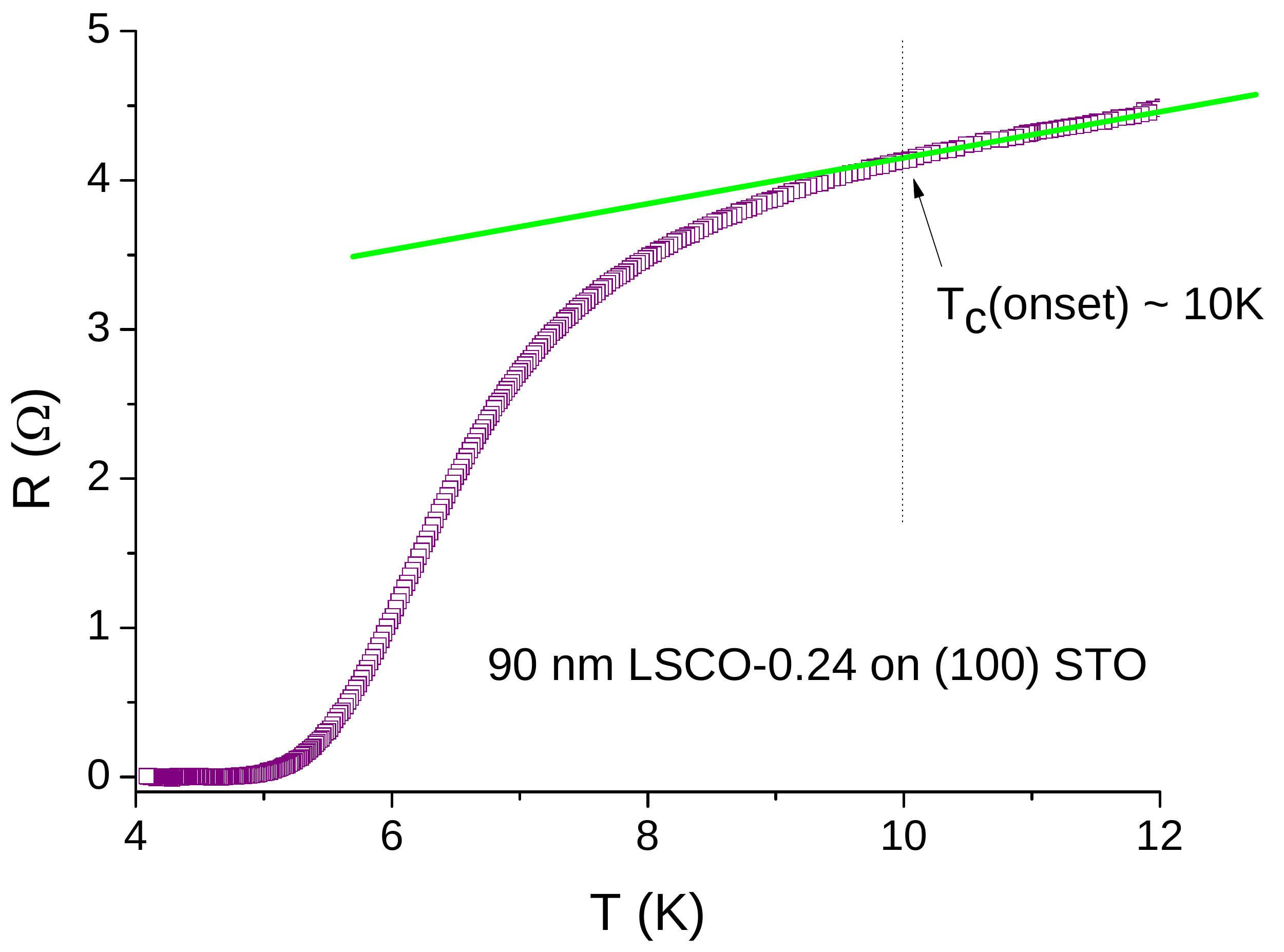}\\
Fig. S1: Resistance versus temperature of a 90 nm thick LSCO-0.24 film on (100) STO wafer.
\end{figure}

We start with the resistance versus temperature shown in Fig. S1 of a bare 90 nm thick overdoped LSCO-0.24 film, laser deposited on (100) $SrTiO_3$ (STO) wafer. One can see that $T_c$(onset) of the deviation from the straight line of the normal resistance is at about 10 K. The transition though is very broad  and reaches zero resistance only at $\sim$5 K. This indicates $Sr$ disorder in the film that is typical of overdoped (and underdoped) cuprate films. There is also a high level of strain in the film on the STO wafer due to lattice mismatch of about 3\% which broadens the transition and lower $T_c$(R=0) even more \cite{strains}. We note that YBCO films on STO are less sensitive to strains, but they are still affected by the Oxygen disorder, more so for the underdoped and overdoped films. We also add here that in our junctions as shown in the top inset to Fig. 1 of the main text, the LSCO films are grown on YBCO, so that the strain effects are much less pronounced.\\

Next, we present in Fig. S2 the resistance versus temperature as measured on two $YBCO/LSCO-0.24/YBCO/Au$
junctions of different barrier thickness on two different wafers. Both junctions show the same $T_c$ of the YBCO electrodes at about 90 K. The junction with the 12 nm thick LSCO-0.24 barrier thickness has a $T_c$(junction) due to the proximity effect in the barrier when it becomes superconducting at about 85 K, while the second junction with the 20 nm thick barrier has a $T_c$(junction) of about 70 K (both are marked with arrows in this figure). This is in line with the stronger pairs decay in the thicker barrier due to the proximity effect, thus allowing for superconductivity to be obtained only at a lower temperature. The resistances below both $T_c$(junction) values are due to the gold leads to the junctions, and their different values is due to different lead lengths.\\

\begin{figure}
\includegraphics[height=9cm,width=13cm]{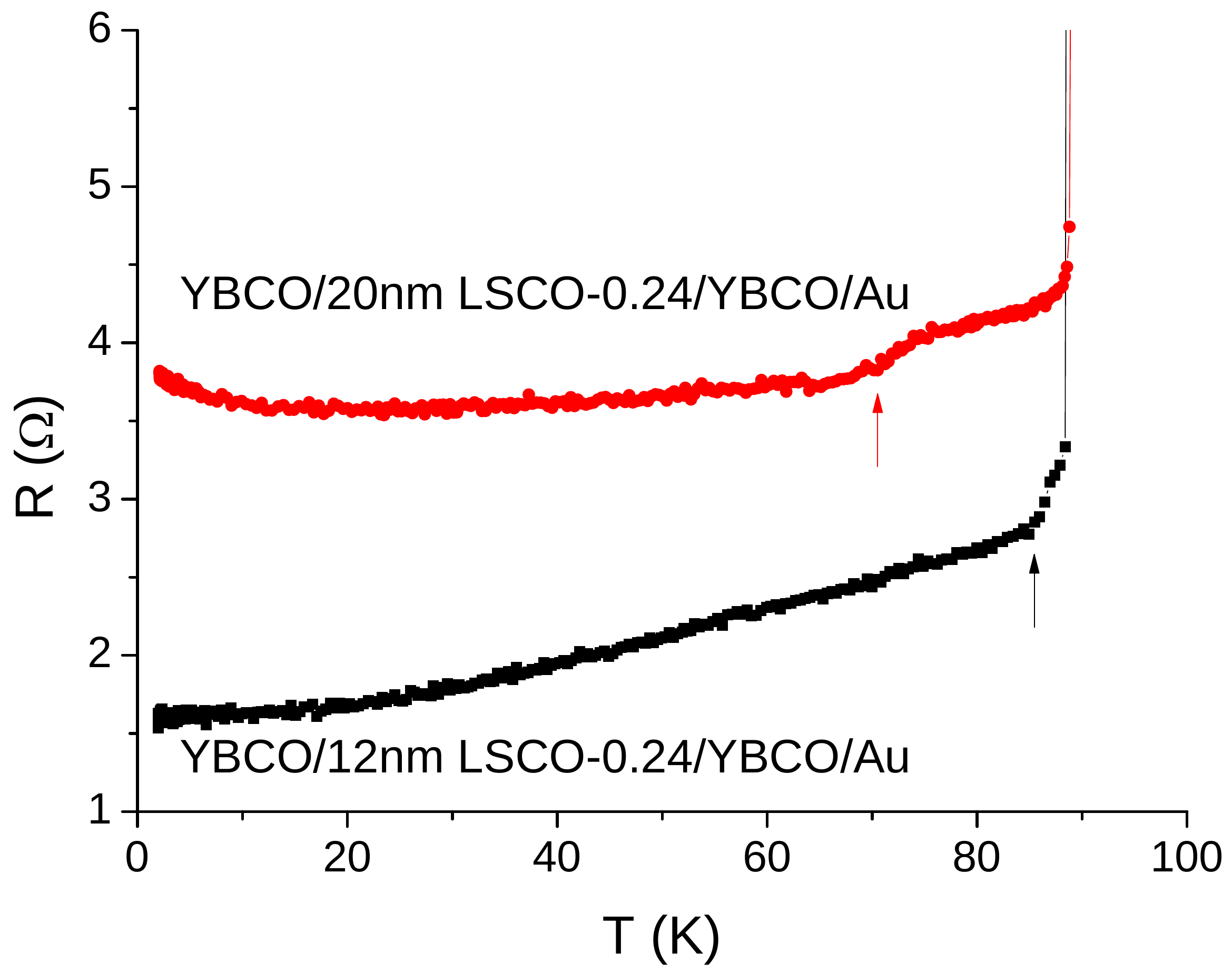}\\
Fig. S2: Resistance versus temperature of two YBCO/LSCO-0.24/YBCO/Au junctions on two different wafers with LSCO-0.24 barrier thicknesses of 12 nm and 20 nm. The transition temperatures $T_c$ of the YBCO electrodes are seen at about 90 K, while the proximity induced transition temperatures of the barriers are marked with arrows. The residual resistance at low temperatures is due to the gold leads to the junctions.
\end{figure}

Figs. S3 and S4 show the dI/dV conductance spectra at various temperatures of the two junctions of Fig. S2. These are the as measured spectra without any vertical shift of the data. As one can see the spectra are mostly "top hat" like, where the horizontal flat lines are due to the gold lead resistance. Once the critical current of the junctions is reached, the conductance goes down, and we marked with arrows on these figures the critical voltages $V_c$ at which this phenomenon occurs. The critical currents $I_c$ can now be determined by the product $V_c \times dI/dV$ where the dI/dV value is taken at the same $V_c$ bias. To improve the statistics of the $I_c$ value determination,  we took the average of both positive and negative values, as marked in the figures. We note that flux-flow at high $I_c$ values makes the drop in conductance at the critical current more rounded, but still one can determine reasonably reliable $V_c$ values as shown here.\\

\begin{figure}
\includegraphics[height=9cm,width=13cm]{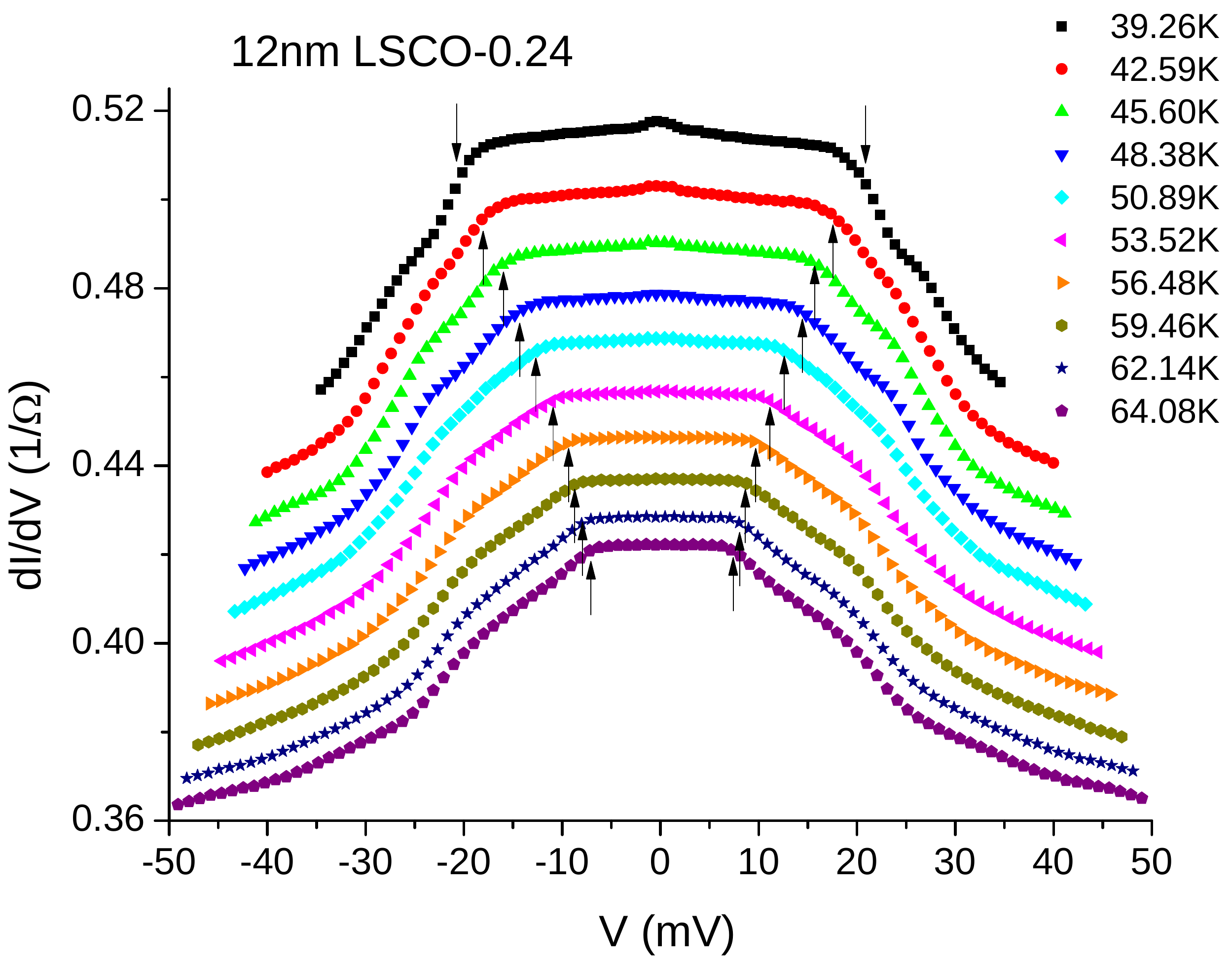}\\
Fig. S3: Conductance spectra of the junction with the 12 nm thick LSCO-0.24 barrier of Fig. S2 at various temperatures. The arrows mark the critical voltages $V_c$ at the critical currents $I_c$ which are equal to $V_c \times dI/dV$ at the same $V_c$. Averages of the $|\pm V_c|$ were taken in the determination of the $I_c$ values.
\end{figure}

\begin{figure}
\includegraphics[height=9cm,width=13cm]{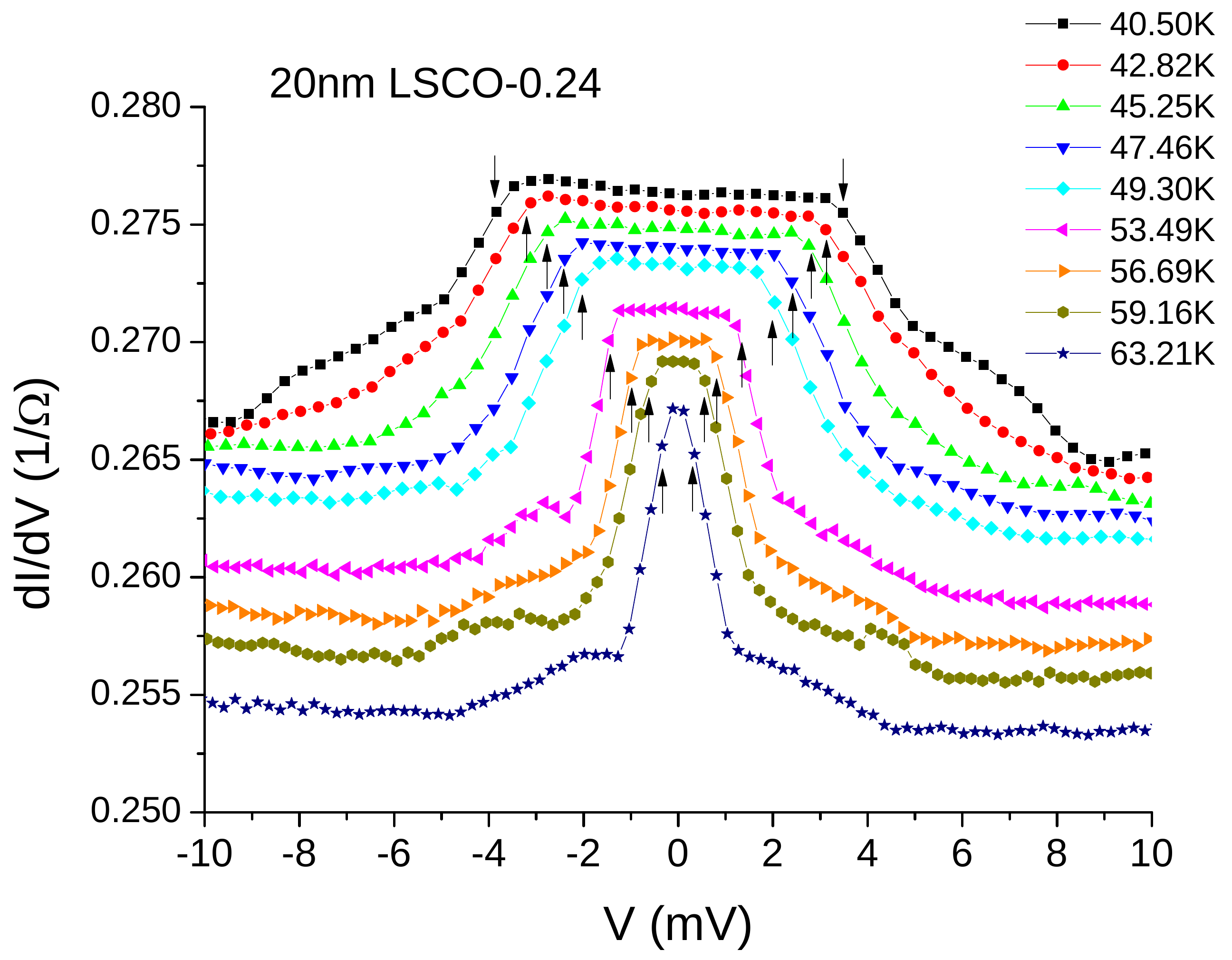}\\
Fig. S4: Conductance spectra of the junction with the 20 nm thick LSCO-0.24 barrier of Fig. S2 at various temperatures. The $I_c$ values were extracted as explained in Fig. S3.
\end{figure}

\begin{figure}
\includegraphics[height=9cm,width=13cm]{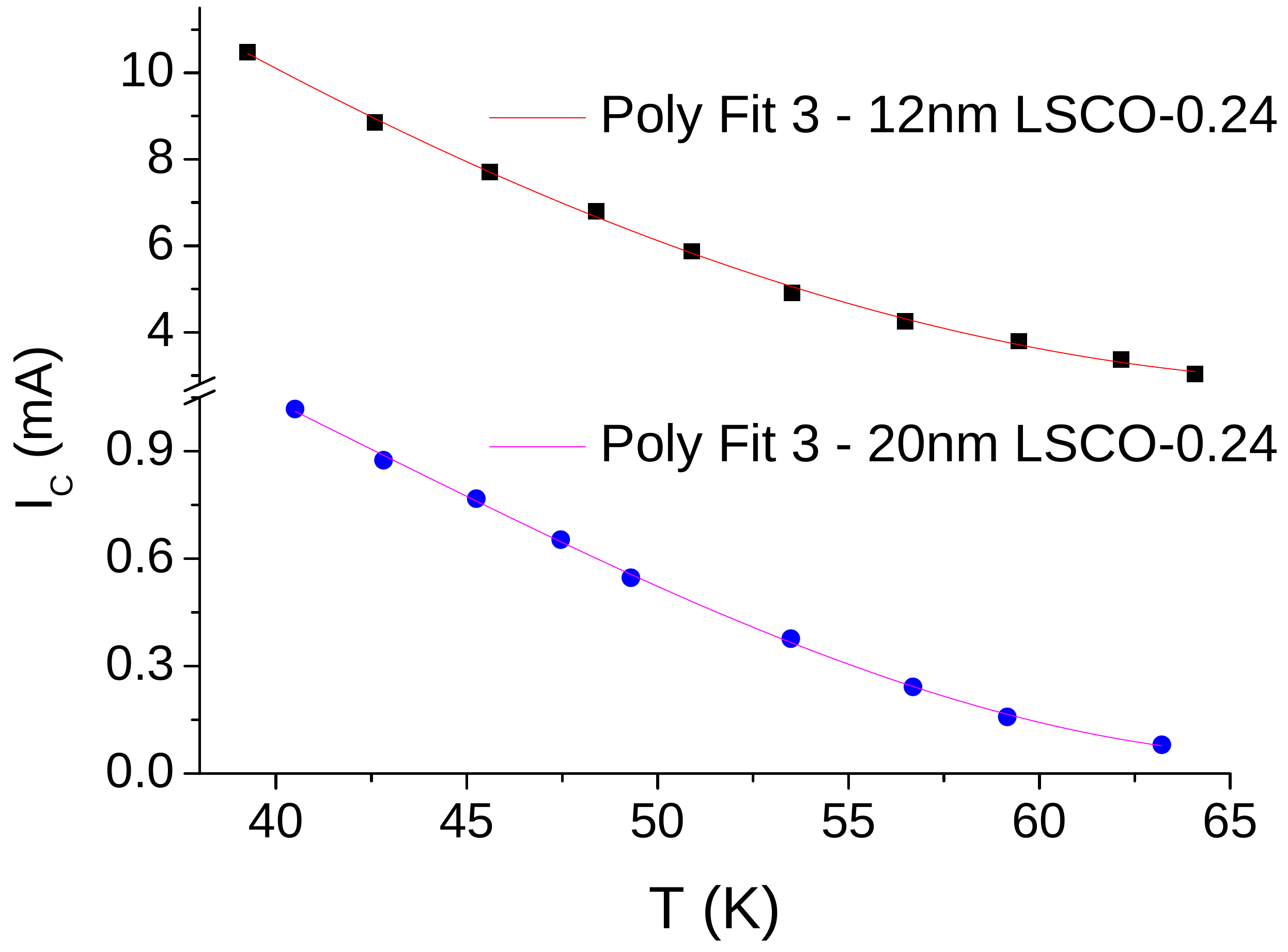}\\
Fig. S5: The critical currents of the two junctions of Figs. S2, S3 and S4, as a function of temperature. The curves are polynomial of order 3 fits to the data.
\end{figure}

The critical currents found from Figs, S3 and S4, are plotted in Fig. S5 as a function of temperature for both junctions, together with polynomial fits of order 3. For any given temperature, there is about an order of magnitude difference in the $I_c$ values of both junctions (note the broken ordinate scale). The fitting curves agree well with the data, and we shall use the interpolated $I_c$ values of one junction when extracting the $\xi_N(T,x)$ values from this data, as explained in the following.\\

\begin{figure}
\includegraphics[height=9cm,width=13cm]{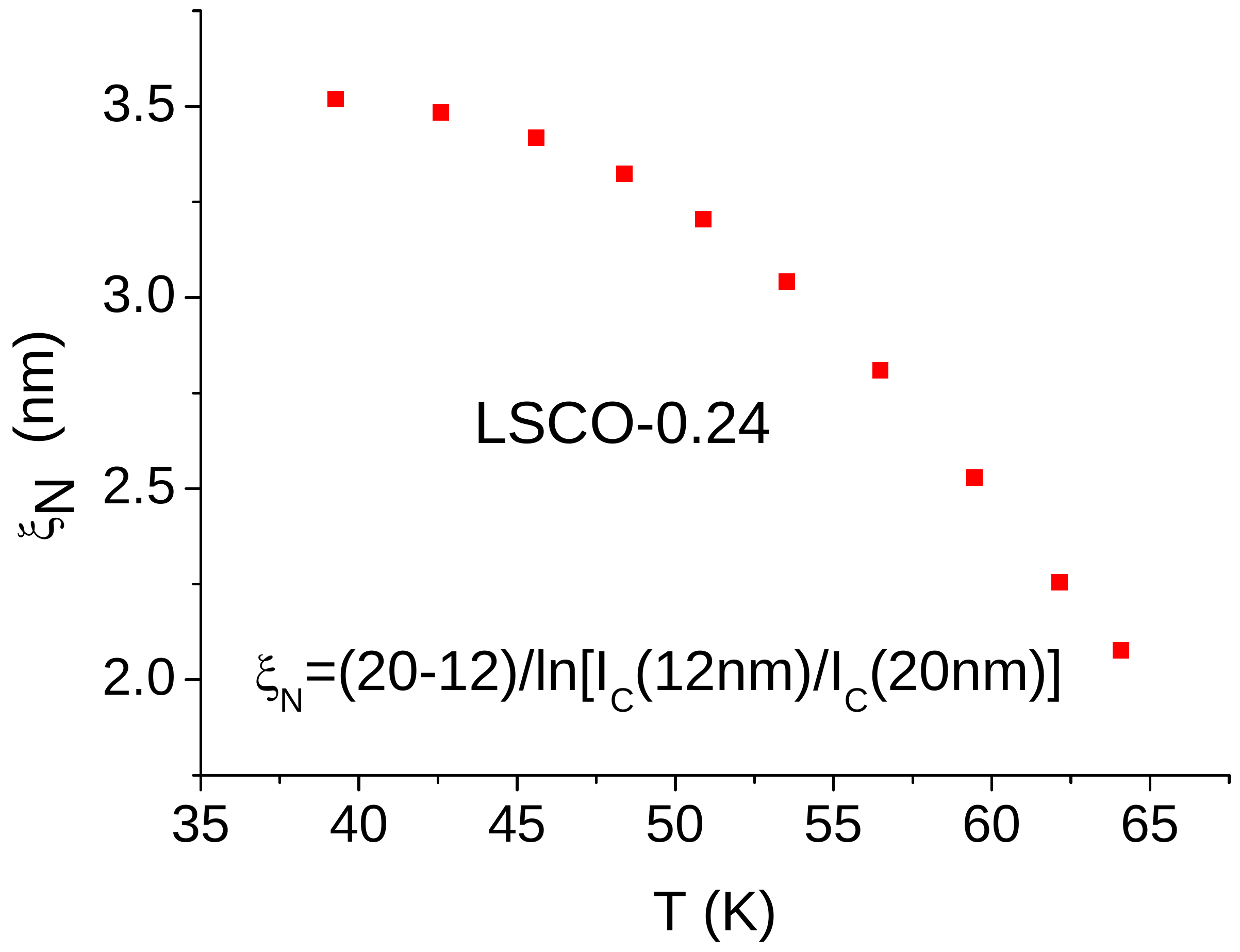}\\
Fig. S6: The normal coherence length $\xi_N$ of LSCO-0.24  as a function of temperature obtained by the formula shown in this figure (Eq. (4)) using the $I_c(T)$ data of Fig. S5.
\end{figure}

\section{Extraction of $\xi_N(T,x)$ from the data}

As said in the main text, we give here a detailed description of how $\xi_N(T)$ is extracted from the $I_c$ data for the LSCO-0.24 junctions as found in Fig. S5. For any given temperature $T$, the standard De Gennes formula for the proximity effect yields \cite{DG}:\\

\begin{equation}I_c = A\, exp[-\xi_N(T)/d]
\end{equation}
where $A$ is a constant and $d$ is the barrier thickness. Fig. S5 allows us to write for each temperature two such equations, one for the junction with the $d$=12 nm thick barrier and the other for the junction with the $d$=20 nm thick barrier. By dividing these two equations, the constant $A$ is canceled out and one finds that:

\begin{equation}\xi_N(T)=\frac{8}{ln I_c(12nm)-ln I_c(20nm)}
\end{equation} 
where 8 is the difference in nm between the two barrier thicknesses. Fig. S6 depicts the resulting $\xi_N(T)$ values versus temperature, which are used in the phase diagram of $\xi_N(T,x)$ in Fig. 5 of the main text, for the $x$=0.24 doping level.\\

\section{Exponential decay of $\xi_N$ versus $d$ in similar junctions }

\begin{figure}
\includegraphics[height=9cm,width=13cm]{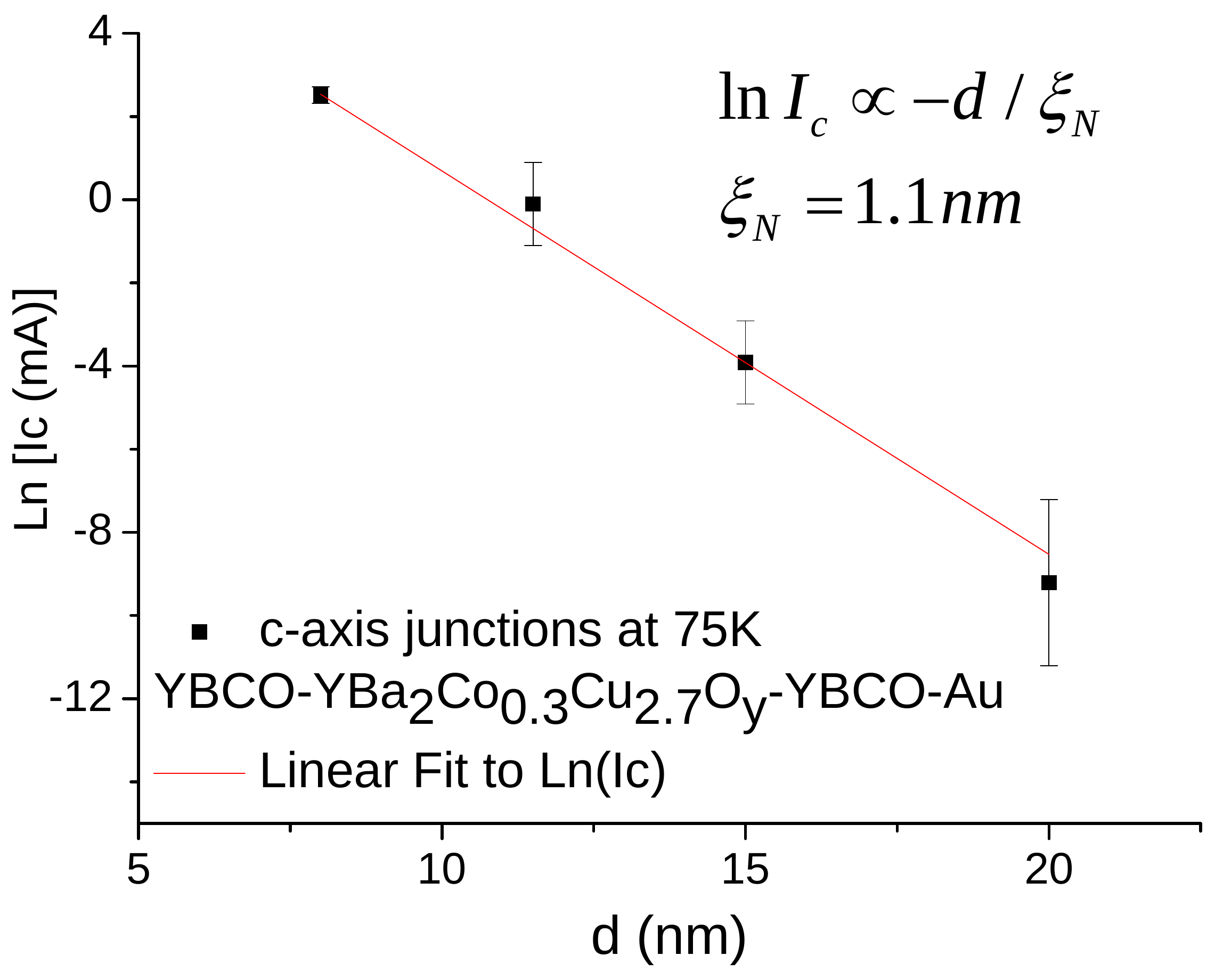}\\
Fig. S7: Critical currents of similar c-axis $YBa_2Cu_3O_{7-\delta}/YBa_2Co_{0.3}Cu_{2.7}O_y/YBa_2Cu_3O_{7-\delta}/Au$ junctions as shown in the top inset to Fig. 1 of the main text, at 75 K versus the barrier thickness $d$. The barrier here is also a cuprate with a lower $T_c$ ($\sim$50 K) which is similar to the LSCO-x barriers in the present study. The good linear fit of the data here on the semi log plot indicates the existence of the standard exponential behavior of the proximity effect. This is in contradiction to the conjectured logarithmic behavior proposed by Marchand et al. for explaning the long range proximity effect \cite{Marchand}.
\end{figure}

Finally, we show experimentally that the standard exponential decay of $I_c$ versus the barrier thickness $d$ as in Eq. (3), is obeyed also in c-axis junctions similar to the present ones, which also exhibit a long range proximity effect. We demonstrate this phenomenon, in order to refute possible logarithmic decays as proposed by Marchand et al. to explain theoretically the long range proximity effect \cite{Marchand}. For this we used $YBa_2Cu_3O_{7-\delta}/YBa_2Co_{0.3}Cu_{2.7}O_y/YBa_2Cu_3O_{7-\delta}/Au$ junctions with the exact geometry as in the top inset to Fig. 1 of the main text, where $YBa_2Co_{0.3}Cu_{2.7}O_y$ is an underdoped cuprate with $T_c$ of about 50 K (instead of the LSCO-x barriers of the present study with $T_c$ of less than 25 K). Measurements of the critical currents $I_c$ at 75 K were done in the same way as described in Figs. S3 and S4, and the data for different $d$ values is shown in Fig. S7. One can see that on the semi-log plot, $ln(I_c)$ decays linearly versus $d$. This demonstrates that the $I_c$ data obeys the conventional exponential decay of the standard proximity effect as given in Eq. (3), with no signs of a logarithmic decay. A $\xi_N$ value of 1.1 nm is deduced from the slope of the linear-fit line of Fig. S7. This value still represents a long range proximity effect, as compared to the expected 0.1-0.2 nm decay length in the c-axis direction \cite{Bozovic}. The reasons why we did not use this barrier in the present study are that first, it is hard to determine its doping level, and second, that it has a narrower range of possible temperatures to work with in the pseudogap regime (above 50 K and below 90 K).\\


\end{document}